\documentclass[iop]{emulateapj}
\usepackage{epsfig,natbib}

\begin{document}

\title{Spectral Eclipse Timing}

\author{Ian Dobbs-Dixon$^{1}$, Eric Agol$^{2,3}$, and Drake Deming$^{3,4}$}

\affil{$^{1}$Department of Physics, NYU Abu Dhabi PO Box 129188 Abu
  Dhabi, UAE\\$^{2}$Department of Astronomy, University of Washington,
  Seattle WA 98195 USA \\$^{3}$NASA Astrobiology Institute Virtual
  Planet Laboratory \\$^{4}$ Department of Astronomy,
  University of Maryland, College Park, MD 20742 USA}

\begin{abstract}
We utilize multi-dimensional simulations of varying equatorial jet
strength to predict wavelength dependent variations in the eclipse
times of gas-giant planets. A displaced hot-spot introduces an
asymmetry in the secondary eclipse light curve that manifests itself as
a measured offset in the timing of the center of eclipse. A
multi-wavelength observation of secondary eclipse, one probing the
timing of barycentric eclipse at short wavelengths and another probing
at longer wavelengths, will reveal the longitudinal displacement of
the hot-spot and break the degeneracy between this effect and that
associated with the asymmetry due to an eccentric orbit. The effect of
time offsets was first explored in the IRAC wavebands by
\citet{williams2006}. Here we improve upon their methodology, extend
to a broad ranges of wavelengths, and demonstrate our technique on a
series of multi-dimensional radiative-hydrodynamical simulations of HD
$209458$b with varying equatorial jet strength and hot-spot
displacement. Simulations with the largest hot-spot displacement
result in timing offsets of up to $100$ seconds in the
infrared. Though we utilize a particular radiative hydrodynamical
model to demonstrate this effect, the technique is model
independent. This technique should allow a much larger survey of
hot-spot displacements with JWST then currently accessible with
time-intensive phase curves, hopefully shedding light on the physical
mechanisms associated with thermal energy advection in irradiated
gas-giants.
\end{abstract}

\section{Introduction}
Highly irradiated, presumably tidally locked, short-period gas-giant
planets have provided invaluable information in our quest to
understand the formation and evolution of exoplanets. Given their
proclivity to transit and their favorable size relative to their host
stars, a wealth of information has been extracted observationally and
they will likely remain the best characterized of all
exoplanets. However, they are significantly different than our own
gas-giant planets. Most striking is the influence of the incident
stellar energy, which provides a luminosity $\sim 10^{5}$ times higher
then the internal luminosity of the planet. The role of this incident
energy in influencing the planet's evolution and shaping the observable
features remains an outstanding question, crucial for interpreting and
understanding the wealth of observations.

A number of groups are studying the atmospheric behavior of gas-giant
planets utilizing multi-dimensional radiative-hydrodynamical
models. For a non-exhaustive list of groups and techniques utilized
see \citet{dobbsdixon2013,heng2014}. Though the details vary between
models, there is a general consensus that highly irradiated, tidally
locked gas-giant planets form broad, supersonic, super-rotating,
equatorial jets. These equatorial jets account for a vast majority of
the dynamical energy transport, and are responsible for differences
between radiative-only calculations and the fluxes actually
observed.

In a very general sense, the efficiency of energy transport via a jet
can be described by two parameters: the radiative time-scale (how long
it takes to cool) and the dynamical time-scale (how long it takes
fluid to move from day to night). \footnote{Note that
  \citet{perezbecker2013} have suggested that though advection is the
        {\it mechanism} for heat transport, the physical process
        governing the efficiency of advection is associated with the
        time-scale for gravity-wave propagation. We are not disputing
        that here, but rather using advection (via the dynamical
        time-scale) to illustrate a general point. Moreover,
        \citet{perezbecker2013} show that in the strong-forcing limit,
        relevant for hot-Jupiters, invoking advective time-scales is
        perfectly valid. For these planets, a comparison between
        $\tau_{rad}$ and $\tau_{adv}$ yields results that agree with
        their numerical simulations.} We expect these two time-scales
to be comparable. For example, consider a highly irradiated planet
where the radiative time-scale is significantly shorter then the
dynamical time-scale; the gas would cool very efficiently as it
travels from day to night. However, the lack of thermal energy
transport to the night-side implies the day-night pressure
differential (the primary driving force for the winds) would be
extremely large. This would shorten the dynamical time-scale, bringing
it closer into line with the radiative time-scale.

As the number of short-period gas-giant planets discovered to be
transiting their host stars continues to increase, we can begin to
make detailed comparisons to predictions from multi-dimensional
radiative-hydrodynamical models. One of the most dramatic set of
observations that demonstrates the importance of dynamics are full or
partial phase curves of short-period planets in the infrared. Though
necessarily hemispherically averaged, it is possible to extract the
longitudinal dependence of the thermal emission across the surface of
the planet \citep{cowan2008_2}. Differences between observed
phase curves and those predicted from a radiative-equilibrium
calculation can be attributed to the transport of thermal energy via
vigorous atmospheric dynamics. Given the current accuracy of
observations, the most telling aspects of a phase curve are the
location of the maximum and minimum flux, and the day-night
temperature differential.

The most accurate phase curve to date is an $8 \micron$ observation of
HD $189733$b by \citet{knutson2007_2,knutson2009,knutson2012}. They
find the peak in the thermal emission from the planet occurs $2.3$
hours before secondary eclipse, indicative of downwind advection by a
strong equatorial jet, and day-night temperature differential of $\sim
240$K. However, offsets and day-night temperature contrasts are not
universal. \citet{cowan2007} monitored the phase-variations of three
planets at $3.6$ and $4.5$ and $8\micron$, finding that only HD
$179949$b exhibited any variation in phase, while observations of HD
$209458$b \citep{zellem2014} and $51$-Peg showed no variation
indicating extremely efficient energy redistribution. It is important
to note that these phase curves are relatively sparsely populated and
consist of non-consecutive observations. The first phase curve of the
non-transiting planet $\nu$-Andromeda
\citep{harrington2006,crossfield2010} suggests a temperature
differential of over $900$K and a hot-spot offset of $\sim
80^{\circ}$.

In this paper we focus on the location of the hot-spot. We currently
do not understand why some planets exhibit hot-spot offsets while
others do not. As discussed above, this should ultimately be related
to variations in dynamical and radiative efficiencies. However, this
simple interpretation may not be sufficient. For example, one might
expect a large hot-spot offset to be correlated to a small temperature
differential. This picture seems to work for HD $189733$b with its
moderate hot-spot offset and several hundred degree temperature
differential. However, the hot-spot on $\nu$-Andromeda is extremely
large and the temperature differential is huge. In hopes of
understanding the importance of composition, incident flux, etc. in
determining the efficiency of heat transfer by the atmosphere,
significantly more data would be extremely helpful. Unfortunately,
full or half orbit phase curves use significant telescope time, and
surveying many systems is difficult. Here we explore a technique for
extracting the longitudinal offset of the hot-spot via a single,
multi-wavelength observation of secondary eclipse utilizing variations
in the time of secondary eclipse. We illustrate the technique by using
multi-dimensional radiative hydrodynamical models of the giant planet
HD $209458$b to predict the timing offsets as a function of
wavelength. In Section (\ref{sec:obs}) we briefly discuss the
multi-dimensional models, explain how to extract emission maps from
the simulations, and discuss the technique for calculating the
wavelength-dependent timing offsets. In Section (\ref{sec:time}) we
present our predictions for HD $209458$b, and discuss the correlation
between timing differences and hot-spot offsets. A variant of this
technique was first proposed by \citet{williams2006}, who used the
simulations of \citet{cooper2005} to predict eclipse-timing offsets
(they refer to these as uniform time offsets) in Spitzer's four IRAC
bandpasses. We compare our methods and results to theirs in Section
(\ref{sec:williams}). Finally, we discuss the feasibility of such a
measurement given current and future instrumental limitations in
Section (\ref{sec:feasibility}). Section (\ref{sec:conclusion})
concludes with a general discussion of our results.

\section{Calculating Observables from 3D Models}
\label{sec:obs}
To illustrate the concept of variations in transit-timing due to an
offset hot-spot we concentrate on simulations of the transiting
gas-giant planet HD $209458$b. We begin by briefly discussing our
simulations, explain the technique for extracting observable
emission maps, and finally the technique used to calculate variations
in transit-timing.

\subsection{Dynamical Simulations}
\label{sec:simulations}
In \citet{dobbsdixon2012_1} we performed extensive simulations of HD
$209458$b utilizing our multi-dimensional radiative-hydrodynamical
model. The model solves the fully compressible Navier-Stokes equations
throughout the entire atmosphere from pressures of $10^{-6}$ to
$500$ bars. The dynamical equations are coupled to a
wavelength dependent stellar energy deposition, radiative-transfer via
flux-limited diffusion, and a two-temperature energy equation. See
\citet{dobbsdixon2012_1} for a discussion on solution techniques,
resolution information, and parameter choices. We compared our results
to observed transit spectra, finding excellent agreement, and
predicted wavelength-dependent timing variations in center of primary
transit. Note that the variations in primary transit timing we discuss
in \citet{dobbsdixon2012_1} differ from what we present here. Though
both related to dynamics, deviations during primary transit are due to
differences in the scale-height of the eastern and western
terminators, while timing variations of the secondary eclipse
discussed here are primarily due to variations in hot-spot location.

To explore the efficiency of the equatorial jet in redistributing
energy throughout the atmosphere, \citet{dobbsdixon2012_1} performed a
number of simulations with viscosity varying from $10^8$ to
$10^{12}\,cm^2/s$. Viscosity serves as a mechanism to
self-consistently convert the kinetic energy of the flow back into
thermal energy and is meant to encapsulate the effect of shocks,
instabilities, or other sources of un-resolved drag on the atmospheric
flow. Large viscosities resulted in slow, sub-sonic flows, no jet
formation, and large day-night temperature differentials. The
atmospheric dynamics in these simulations did little to change the
radiative temperature distribution and the hottest point on the planet
remained near the sub-stellar point. However, as we decreased the
viscosity, we found that a super-rotating, supersonic, equatorial jets
formed that were able to efficiently transport energy to the
night-side while simultaneously advecting the hot-spot downwind, east
of the sub-stellar point. The smaller the viscosity, the larger the
offset. Figure (3) of \citet{dobbsdixon2012_1} shows the temperature
and zonal velocity distribution at the photosphere of the planet for a
range of viscosities. Although we utilize these simulations to
illustrate this phenomena, we emphasize that the techniques we present
are model independent and can be applied to any observation of
secondary eclipse.

\subsection{Observables}
\label{sec:calc_obs}
Multi-dimensional radiative-hydrodynamical calculations provide us
with a three-dimensional distribution of temperature and density
throughout the atmosphere that we can utilize to explore observational
phenomena. Given temperature and density, detailed gas opacities from
\cite{sharp2007} allow us to calculate wavelength-dependent emission
and absorption spectra, and phase curves. Of primary interest to us in
this paper, are the wavelength-dependent maps of emission from the
day-side of the planet. Two components contribute to this emission:
thermal emission and scattered light. We discuss each in turn below.

\subsubsection{Thermal Emission}
To calculate the thermal emission from the day-side of the planet we
integrate inwards along rays parallel to the line of sight. The
positional dependent intensity required to calculate the detailed
secondary eclipse light curve is given by
\begin{equation}
  I_{emis}\left(x,y,\lambda\right) = \frac{\lambda^2}{c} \int_{0}^{\infty}
  B_{\lambda}\left(x,y,\tau\right)e^{-\tau_{\lambda}}d\tau_{\lambda}.
  \label{eq:intensity}
\end{equation}
We assume the atmospheric gas emits as a Blackbody, given by
\begin{equation}
 B_{\lambda}\left(x,y,\tau\right) = \frac{2\pi
   hc^2/\lambda^5}{exp\left(\frac{hc}{\lambda
     kT}\right)-1}.
 \label{eq:emerging}
\end{equation}
The density and temperature values, used in Equation
(\ref{eq:emerging}) and needed to calculate $\tau$ at each location,
are interpolated from the values calculated in the 3D models. Finally
the total apparent day-side luminosity can be obtained by integrating
over the observed disk
\begin{equation}
  L_p\left(\lambda\right) = \int I_{p}dA,
\end{equation}
where $dA$ is the area element in the observer's plane.

\subsubsection{Scattered Light}
We assume a uniform albedo and diffuse scattering for the planet.
To calculate the incident stellar light scattered from the planet we
first consider the specific intensity emitted in the direction of the
observer. The frequency dependent intensity scattered from each
longitude ($\phi$) and latitude ($\theta$) on the planet can be
written as
\begin{eqnarray}
I_{scat}\left(\phi,\theta,\nu\right) & = &
A_S\left(\phi,\theta,\nu\right)I_{inc}cos\phi cos\theta \\ \nonumber
 & = & \left(\frac{3}{2}\right) A_g\left(\phi,\theta,\nu\right)I_{inc}cos\phi
cos\theta,
\label{eq:iscat}
\end{eqnarray}
where $A_S$ and $A_g$ are the spherical and geometric albedo
respectively. The scattered intensity is understood to only be
relevant for the longitudes and latitudes both exposed to the
stellar light, and visible to the observer during the eclipse;
{\it i.e.} within $\phi,\theta=\pm\pi/2$ of the sub-stellar point
$\left(\phi,\theta\right)=\left(0^{\circ},0^{\circ}\right)$. The
$cos\phi cos\theta$ term accounts for the angle between the local
normal and the incident stellar intensity,
$I_{inc}=B_{\star}\left(\nu\right)\left(\frac{R_{\star}}{a}\right)^2$.

The light scattered toward the observer can be calculated by
integrating Equation (\ref{eq:iscat}) over the entire observable
hemisphere, adding a geometric term to account for the reduced area
visible and the standard spherical area element. This gives the
intensity scattered towards the observer,
\begin{equation}
I_{scat}\left(\nu\right) = R_p^2 \left(\nu\right) \int \int
I_{scat}\left(\phi,\theta,\nu\right) cos\phi_{obs}
cos^2\theta_{obs}d\phi d\theta.
\label{eq:iscat_tot}
\end{equation}
In the limit of spatially constant albedo, this can be integrated and
written in the usual form as
\begin{equation}
I_{scat}\left(\nu\right) = R_p^2 A_g\left(\nu\right)
I_{inc}\left(\nu\right)
\frac{sin\alpha+\left(\pi-\alpha\right)cos\alpha}{\pi}.
\label{eq:iscat_analytic}
\end{equation}
The final fraction is the well known Lambertian phase-function. In the
following we assume a spatially constant albedo. However, theoretical
calculations \citep{lee2015} and multiple observations
\citep{demory2013,esteves2014,hu2015,shporer2015} suggest that cloud
coverage may not be spatially uniform. In this case, one must use
Equation (\ref{eq:iscat_tot}) for the scattered light contribution.

\subsubsection{Calculating Timing Offsets}
Utilizing the total surface brightness of the planet, we generate
simulated secondary eclipse light curves by integrating over the
visible portion of the planet at each time step.  We used the planet
radius and orbital parameters derived in \citet{torres2008}. Once
wavelength dependent simulated light curves are computed, we utilize
the standard observational technique to fit them using a secondary
eclipse model that {\it assumes} a uniform planet surface brightness
\citep{mandel2002}. If the planet was in radiative equilibrium the
mid-eclipse time of transit would coincide with the midpoint of
barycentric secondary eclipse. Deviation in the planet's surface
brightness gives the so-called ``uniform time offset"
\citep{williams2006}; this corresponds to the artificial time inferred
using the wrong ({\it i.e.} uniform) planet emission model.

\section{Predicted Timing Offsets}
\label{sec:time}
Utilizing the simulations of HD $209458$b from
\citet{dobbsdixon2012_1} and the techniques described in Section
(\ref{sec:obs}) we calculate the variation in the time of central
eclipse. Figure (\ref{fig:timing}) illustrates our results for models
with varying viscosity and jet strength using the both the thermal
emission and scattered light, $I_{tot} = I_{emis} + I_{scat}$. As
expected, smaller viscosities and stronger equatorial jets are
associated with larger timing offsets. For simulations with large
hot-spot displacements we predict timing offsets of up to $96$ seconds
at $2\micron$. At longer wavelengths Figure (\ref{fig:timing}) shows a
general trend of increasing timing offset with decreasing
wavelength. This is the result of increasingly spatially inhomogeneous
(patchy) emission at shorter wavelengths (see Figure
\ref{fig:imap}). Due to the steep slope of the emission spectra in the
Wien tail, small changes in gas temperature lead to significant
changes in emission resulting in patchy emission and large timing
offsets. This effect becomes more pronounced at shorter
wavelengths. However the scattered light, inherently symmetric based
on our assumptions, acts to damp out any timing offsets seen at even
shorter wavelengths. For calculations assuming a small albedo ($0.05$)
this cutoff is approximately $1\micron$, while those assuming a larger
albedo ($0.35$) the cutoff wavelength is somewhat longer, at
$3\micron$. Note that though we include the effects of albedo on
timing predictions, we have not self-consistently modified the stellar
energy deposition in the multi-dimensional simulations. All
simulations assume zero albedo, while a non-zero albedo is used in
post-processing the simulated atmospheres with our radiative transfer
code. Thus, a small error in the temperature of the planet is made of
the order $(1-A)^{1/4}$; this will affect our results quantitatively,
but we expect will capture the same qualitative features as a fully
self-consistent computation. The detailed features seen in all curves
in Figure (\ref{fig:timing}) are associated with peaks in the opacity,
primarily due to water. A larger opacity implies we are probing higher
in the atmosphere where the planet asymmetries are larger, while
wavelengths corresponding to smaller opacities probe deeper, more
symmetric regions of the atmosphere, resulting in smaller time
offsets. In Table ({\ref{table:one}) we detail eclipse timing offsets
  averaged over J, H, K, IRAC, and proposed James Webb Space Telescope
  (JWST) bandpasses.

\begin{figure}
  \centering
  \includegraphics[width=1\linewidth]{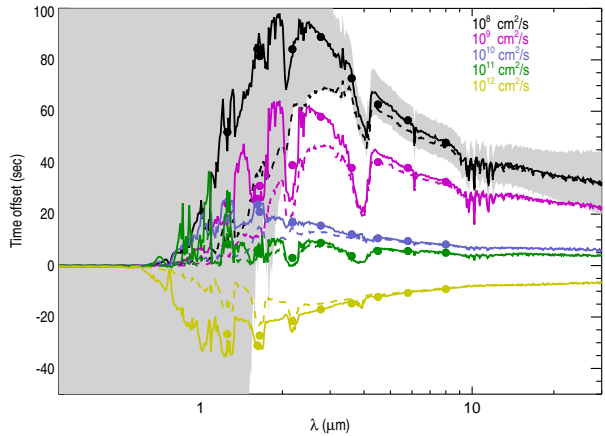}
  \caption{Variation in the time of secondary eclipse for models of HD
    $209458$b with varying viscosities and jet strengths. The solid
    lines are calculated assuming an albedo of $0.05$ while the dashed
    lines assume a value of $0.35$. Solid dots show the band-averaged
    time offsets reported in Table (\ref{table:one}). The shaded
    region shows the extent of the observational errors on the time
    offset for a planet with a viscosity of $10^8\,cm^2/s$ around
    sun-like star at $10$ parsecs viewed by JWST (See discussion in
    Section (\ref{sec:feasibility})).}
  \label{fig:timing}
\end{figure}

To further illustrate the relative contributions of scattered light
and thermal emission we show intensity maps from the day-side of the
planet for a range of wavelengths in Figure (\ref{fig:imap}). The top
row shows the contribution from thermal emission, the middle from
scattered light, and the bottom shows the total. Thermal emission maps
are calculated using the lowest viscosity simulation ($10^8\,cm^2/s$) while
scattered light maps assume an albedo of $0.05$. At
$\lambda=0.7\micron$ (left column) the peak of the thermal emission
exhibits a large offset, but is orders of magnitude smaller then the
scattered light and is completely masked in the total emission. At
$\lambda=10\micron$ (right column) the thermal emission is also
offset, though not as strongly peaked, and several orders of magnitude
brighter then the scattered light. The total emission is thus
dominated by the thermal component. At intermediate wavelengths
($\lambda=1.3\micron$, middle column) the thermal and scattered
intensities are comparable and the resulting peak of the total
emission is pulled back toward the sub-stellar point.

\begin{figure}
  \centering
  \includegraphics[width=1.0\linewidth]{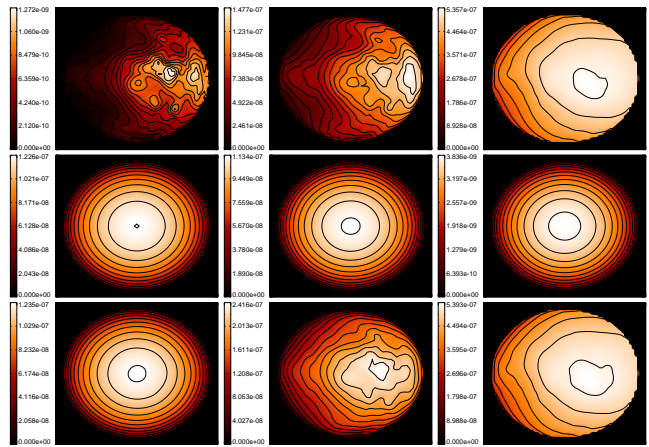}
  \caption{The day-side intensity from thermal emission (top row), the
    scattered intensity (middle row), and the total intensity (bottom
    row) at several representative wavelengths. The first column shows
    the results for $\lambda=0.7\micron$, the middle column shows
    $\lambda=1.3\micron$, and the final column shows
    $\lambda=10\micron$. The sub-stellar point is located at the center
    of each image. Note the scale of each plot is different.}
  \label{fig:imap}
\end{figure}

\begin{table}[ht]
\begin{center}
\begin{tabular}{|c|c|c|c|c|c|c|}
\hline
Band & $10^{8}$ & $10^{9}$ & $10^{10}$ & $10^{11}$ & $10^{12}$ & Williams \\
\hline
J     & 51.9 & 24.0 & 18.4 & 8.4 & -26.7 & - \\
H     & 81.4 & 31.0 & 20.8 & 4.8 & -27.2 & - \\
K$_s$ & 84.1 & 39.0 & 16.7 & 3.0 & -21.6 & - \\
F162M & 84.2 & 25.9 & 23.7 & 6.0 & -31.0 & - \\
F277W & 88.9 & 57.9 & 15.5 & 8.8 & -17.0 & - \\
IRAC1 & 72.8 & 39.0 & 12.1 & 3.7 & -14.7 & 53.0 \\
IRAC2 & 62.6 & 40.2 & 10.6 & 5.7 & -12.2 & 49.5 \\
IRAC3 & 56.5 & 38.0 & 9.5 & 5.7 & -10.6 & 44.0 \\
IRAC4 & 47.8 & 32.5 & 8.2 & 5.0 & -9.2 & 37.5 \\
\hline
\end{tabular}
\end{center}
\caption{Band averaged eclipse time offsets in seconds for several
  representative bandpasses (including the proposed F162M and F277W
  JWST filters) for models with viscosities ranging from $10^8$ to
  $10^{12}\,cm^2/s$. The final column lists results from our
  $\nu=10^8\,cm^2/s$ simulation using the methodology of
  \citet{williams2006}.}
\label{table:one}
\end{table}

\subsection{Comparison with \citet{williams2006}}
\label{sec:williams}
Studying the effect of an offset hot-spot on the time of center of
eclipse was first proposed by \citet{williams2006}. Referring to this
phenomena as a uniform time offset, they predicted signals of $86$,
$77$, $67$, and $57$ seconds for the IRAC wavebands $1-4$ and explored
the probability of such a detection. Our current predictions differ
from \citet{williams2006} in three important ways: the underlying
radiative-hydrodynamical method and solution, the method for
calculating intensity maps, and the wavelength range explored. Though
the last point is most crucial, we explore each below in turn.

The first difference between the predictions of \citet{williams2006}
and that presented here is that the underlying multidimensional
radiative-hydrodynamical model used to generate predictions is
different. They utilize results from the model of \citet{cooper2005},
which is a general circulation model (GCM) solving the primitive
equations coupled to a simplified Newtonian cooling
scheme. \citet{cooper2005} predicts an hot-spot offset of $60^{\circ}$
in longitude at the wavelength-averaged photosphere. Our simulations,
described in detail in Section (\ref{sec:simulations}) utilizes a
significantly different numerical code and can produce a range of
hot-spot offsets associated with varying viscosity.

The second difference is the manner of producing the intensity map. In
\citet{williams2006} they utilized the calculations of
\citet{fortney2005} to identify approximate pressures that correspond
to the photosphere of the planet when viewed in each of the four IRAC
bandpasses ($105$, $50$, $35$, and $24$ mbar for IRAC1 through
IRAC4). They then identify the corresponding temperature at each
latitude/longitude, assume that each patch of the planet emits as a
blackbody, and sum up all the contributions to get the total thermal
emission. In this work, as described in Section (\ref{sec:calc_obs}),
we integrate the emission contributions along the line of sight, thus
accounting for the fact that the path of the light is increasingly
non-radial as you move toward the limb of the planet. The emitted flux
does not simply come from a single pressure, but rather a range around
the $\tau=1$ surface associated with the contribution function
\citep{chamberlain1987,knutson2009} as you integrate along the
path. We additionally include the important contribution from
scattered light. Though not relevant at the longer wavelength IRAC
bands, it is imperative to include scattered light at wavelengths near
and below $\sim 1\,\mu m$. The initial assumption in the community at
large was that optical secondary eclipses would never be observable
(likely explaining why Williams did not include it). However,
particularly for very hot planets such as Wasp-12b \citep{hebb2009},
we find that including scattered light is necessary below $\sim 1\,\mu
m$. The final difference is that we explore the eclipse timing offsets
over a much wider range of wavelengths. Though this may appear
trivial, we suggest that the actual measurement of this effect will
require a differential measurement between two or more
wavelengths. The barycentric eclipse cannot be determined from radial
velocities to sufficient accuracy to determine the offset from a
single wavelength. By expanding our wavelength coverage, observers may
choose two wavelength regions with maximal timing differentials,
yielding a potentially observable signature.

To test the importance of the technique used when producing an
emission map of the planet, we produced IRAC light curves utilizing
the blackbody radiation associated with the constant pressure surfaces
of $105$, $50$, $35$, and $24$ mbar (following \citet{williams2006}
technique) utilizing our lowest viscosity ($10^8\,cm^2/s$)
simulation. These pressures are taken from \citet{fortney2005} as the
mean pressures associated with the IRAC photospheres. Predictions
associated with constant pressure surfaces are listed in the last
column of Table (\ref{table:one}) and should be compared to values in
the first column. Though there is reasonably good agreement, our
non-radial ray-tracing method of calculating thermal emission maps
yields timing offsets that are systematically $10-20$ seconds larger
then when utilizing the constant pressure method. This can be
understood by considering where in the planet the observed flux is
actually coming from. As you move away from the sub-observer point the
contribution function ({\it i.e.} the region where photons originate)
moves to lower and lower pressures. Put another way, when integrating
a photons path backwards along the correct slant-geometry (as opposed
to a path normal to the planet's surface) you will reach the
$\tau\left(\lambda\right)=1$ surface much more rapidly. Therefore, as
you near the limb of the planet you are looking further up in the
atmosphere where dynamical redistribution has an increasingly larger
roll. We did not extend the constant-pressure analysis to shorter
wavelengths as short-ward of $\sim 1\,\mu m$ the methods disagree
substantially due to our added contribution of scattered light. It is
important to emphasize that we cannot directly compare our results to
those in \citet{williams2006} as the underlying
radiative-hydrodynamical models and thus the pressure-temperature
distributions are different. Our calculations of constant pressure
maps are derived from the simulations from \citet{dobbsdixon2012_1}.

\subsection{Observational Feasibility}
\label{sec:feasibility}

To explore the feasibility of measuring a secondary eclipse time
offset due to the effects of atmospheric circulation we use numerical
eclipse curves to calculate the photometric precision needed to
achieve a given timing error. We construct two eclipse curves using
the \citet{mandel2002} method, normalizing them to unity during
eclipse (star only). One eclipse curve represents the barycentric
eclipse and the other is shifted in time by $t$ seconds. We subtract
these two eclipse curves, and calculate the standard deviation of the
photometric difference. That standard deviation represents the
photometric error level necessary to detect a shift of $t$ seconds, to
$1-\sigma$ precision. We checked our calculation by comparing to the
analytic formulae given by \citet{carter2008}, (their Eq. 23) finding
agreement to about 3-percent. That level of agreement is consistent
with the difference between our adopted eclipse curve shape
\citet{mandel2002}, and the trapezoidal shape illustrated in Figure 1
of \citet{carter2008}.

To calculate the observed photometric error level in comparison to the
required error level defined above, we assumed that Poisson-counting
noise dominates the photometric uncertainties. Indeed, observations
with Spitzer have achieved close to shot-noise precision
\citep[e.g.]{deming2014}. We used ATLAS stellar models
\citep{castelli2003} in order to calculate the photon flux at JWST,
and we adopted a telescope area of $25m^2$, a throughput (electrons out
divided by photons in) of $20\%$, and a filter bandwidth of $20\%$.

In Figure (\ref{fig:snr}) we show the S/N for the timing offset
itself, as a function of wavelength, based on the modeled timing
offsets from Figure (\ref{fig:timing}), for a planet eclipsed by a
Sun-like star at $20$ parsecs. The shape of this curve is a confluence
of the number of photons from the planet (as per above) and the size
of the timing offset. Values of timing S/N peak at $8.4$ at
$5\micron$, feasibly detectable by JWST, particularly when combining
multiple observations. We show the expected errors associated with
this S/N in Figure (\ref{fig:timing}) as a shaded region for the model
with a viscosity of $10^8\,cm^2/s$. Though the errors are enormous at
wavelengths less than approximately $2\micron$ (primarily due to a
lack of photons) they are reasonable at longer wavelengths. Note that
the curve in Figure (\ref{fig:timing}) is a theoretical prediction,
thus independent of observational resolution. Actual observations will
result in substantially lower resolution. Once a particular
observation has been made, the theoretical data can be appropriately
binned for comparison.

\begin{figure}
  \centering \includegraphics[width=1.0\linewidth]{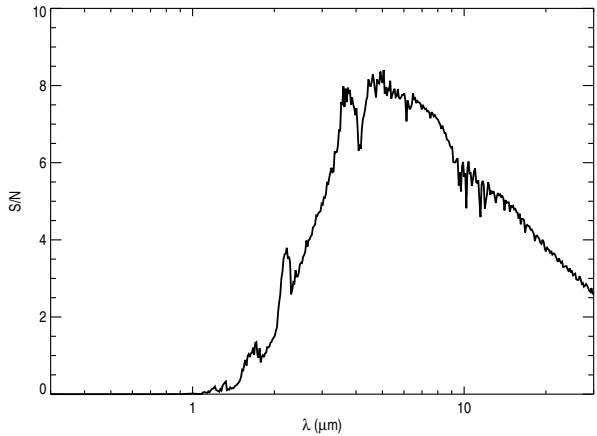}
  \caption{The signal-to-noise as a function of wavelength for
    detection of the dynamically induced timing offsets for a short
    period planet around a sun-like star at $20$ parsecs assuming a
    $20\%$ throughput on JWST and a $20\%$ bandpass. Timing offsets
    are taken from Figure(\ref{fig:timing}) for the lowest viscosity
    simulation and assume an albedo of $0.05$.}
  \label{fig:snr}
\end{figure}

\section{Conclusions}
\label{sec:conclusion}
In this paper we propose a method to determine the horizontal
displacement of the hot-spot on a tidally locked, highly irradiated,
gas-giant planet. A hot-spot displaced from the sub-stellar point will
introduce an asymmetry in the secondary eclipse light-curve that, when
fit utilizing a model that assumes a symmetric planet will manifest
itself as a timing offset of the calculated center of eclipse. In
principle, this measurement could be done utilizing a single
multi-wavelength observation of secondary eclipse. It is crucial to
observe the eclipse in multiple wavelengths to compare calculated
timing offsets to break the degeneracy associated with an eccentric
orbit, as the predicted center of barycentric transit derived from
radial velocity curves is not known to sufficient accuracy.

Though the method proposed in this paper is model independent we
utilized radiative-hydrodynamical simulations from
\citet{dobbsdixon2012_1} of HD $209458$b to illustrate this method
further and predict timing offsets. Simulations develop super-rotating
equatorial jets that redistribute energy from day to night and
displace the hot-spot from the sub-stellar
point. \citet{dobbsdixon2012_1} ran simulations for a range of
viscosities, resulting in varying jet strengths and advective
efficiencies. The hot-spots in simulations that develop little or no
circumplanetary jet remain near the sub-stellar point, resulting in
very little predicted time offset. Conversely, in simulations with
strong, supersonic equatorial jets, the hot-spot is significantly
displaced and results in timing offsets up to $\sim 100$ seconds at
$2\micron$.

The concept of an offset hot-spot inducing a shift in the time of
eclipse was first proposed by \citet{williams2006}. In this paper we
have built upon their work in four important ways: improving the
underlying radiative-hydrodynamical models, modifying the method of
producing thermal emission maps, accounting for the scattered light
contribution, and extending the calculation to a large wavelength
range. The extension of the technique to multi-wavelength is perhaps
the most crucial. When viewed in a single channel the effects that
\citet{williams2006} proposed can be interpreted as either an
eccentric orbit {\it or} non-uniform planetary emission; by itself a
single measurement is not conclusive. A timing differential between
two or more wavebands will break the degeneracy between the eccentric
and asymmetric interpretations.

There have been a number of attempts to measure the time offsets for
planets at multiple wavelengths, though not necessarily
simultaneous. Observations of HD $209458$b by \citet{knutson2008} in
all four Spitzer-IRAC wavebands suggests time offsets of up to
$-12.6\,\pm\,3.5$ minutes for the $4.5\micron$ channel and
$2.0\,\pm\,3.1$ for the $8\micron$ channel. However, our models
predict a maximum differential between these wavebands of only $14.8$
{\it seconds}, so the observed $14.6\,\pm\,4.7$ minute differential is
{\it significantly} larger. Measurements of time offsets for HD
$209458$b by \citet{crossfield2012} in the $24\micron$ MIPS bandpass
suggest a $32\,\pm\,129$ second offset, more in line with our
predictions. Re-examination of the data for HD $209458$b by
\citet{dimondlowe2014} suggests that the observational technique used
previously resulted in much larger errors then originally
quoted. Similarly, IRAC observations of of HD $189733$b utilizing a
technique similar to \citet{knutson2008} suggested differential timing
offsets on the order of minutes, reaching $5.6\,\pm\,0.8$ minutes in
the $3.6\micron$ channel \citep{charbonneau2008}. However,
\citet{agol2010} used a combination of $14$ eclipses of HD $189733$b
at $8\micron$ to derive a $38\,\pm\,11$ second unaccounted for shift
in the center of eclipse that they attribute to an offset
hot-spot. Though only measured in one waveband, this measurement is
much more in line with our predictions from models with a viscosity of
$10^8\,cm^2/s$. With these varying differential measurements and the
associated uncertainty in observing strategies, additional
observations exploring the multiwavelength timing offsets are
necessary.

Ultimately, the method have described is a fairly easy way to extract
information on the dynamically induced shift in the hottest point on
the exoplanet. A multiwavelength observation with an instrument such
as NIRSpec on JWST will allow us to differentiate between models with
varying jet speed, giving us a hint of the underlying physical
phenomena operating in the atmosphere. There are however, more
detailed approaches that utilize the shape of the secondary eclipse to
'map' the disk of the planet \citep{rauscher2007, majeau2012,
  dewit2012}. The measurement described in this paper is complementary
to these more detailed approaches; if measurements of time offsets
indicates a significant wavelength dependent variation, one can
attempt a secondary eclipse map if the S/N warrants it.

In conclusion, the location of the hot-spot is one of the most
striking examples of the importance of dynamics in the atmospheres of
irradiated planets. The precise location will offer clues to the
interplay between the radiative and dynamical processes. Determining
the strength of the offset across a diverse family of planets should
help us to understand the physical phenomena involved. Observations of
full or half orbit phase curves of planets on $\sim 3$ day orbits are
extremely valuable, but also use a significant fraction of telescope
time. The direct correlation between hot-spot displacement and the
wavelength dependent offset of center of secondary eclipse provides
another method of detecting the displacement. Simultaneously
monitoring the timing offset during a single secondary eclipse in
several wavelengths is a much more efficient technique and our
analysis indicates such a measurement will be feasible with JWST. This
more efficient technique would allow for a much larger survey studying
the connection between star, planet, and orbital properties and the
strength of the dynamics.

\section*{Acknowledgments}
 We thank Nick Cowan for valuable discussions and Adam Burrows for
 providing opacities. This work was partly performed as part of the
 NASA Astrobiology Institute's Virtual Planetary Laboratory Lead Team,
 supported by the National Aeronautics and Space Administration
 through the NASA Astrobiology Institute under Cooperative Agreement
 solicitation NNH05ZDA001C. Additional support for this work was
 provided by NASA through grant number 12181 from the Space Telescope
 Science Institute, which is operated by AURA, Inc., under NASA
 contract NAS 5-26555. We would also like to acknowledge the use of
 NASA's High End Computing Program computer systems. Additional
 support for this work was provided by NASA through an award issued by
 JPL/Caltech. We acknowledge support from NSF CAREER Grant
 AST-0645416.

\bibliographystyle{mn2e}
\bibliography{ian}

\end{document}